\documentclass[twocolumn, secnumarabic, amsmath, amssymb, showkeys, nobibnotes, aps, pra, reprint, superscriptaddress]{revtex4-2}

\usepackage{graphicx}
\usepackage{dcolumn}
\usepackage{bm}
\usepackage{booktabs}

\usepackage{upgreek}
\usepackage{amsmath,amssymb}
\usepackage{braket}
\usepackage{xcolor}
\usepackage{ulem}
\usepackage{soul}
\usepackage[colorlinks=true, linkcolor=blue, urlcolor=blue]{hyperref}

\begin{document}

\preprint{}

\title{Quantum key distribution over a 2 km free-space channel with a high secure key rate}

\author{Kyungdeuk Park}
\email{kpark@add.re.kr}
\author{Dongkyu Kim}
\author{Dong-Gil Im}
\author{Yonggi Jo}
\author{Jisu Kim}
\author{Jonguk Choi}
\author{Yong Sup Ihn}
\email{yong0862@add.re.kr}
\affiliation{Agency for Defense Development, Daejeon 34186, Korea}
\date{\today}

\begin{abstract}
  Free-space quantum key distribution (QKD) provides crucial advantages, including mobility and deployment flexibility, for securing next-generation communication networks.
However, atmospheric turbulence and beam wander severely degrade single-photon coupling efficiency and link stability over outdoor channels.
In this work, we experimentally demonstrate a free-space decoy-state BB84 QKD system operating at 803 nm with a 100 MHz repetition rate over a 2 km outdoor link.
By employing an active beam-wander correction based on fast-steering mirrors (FSMs) and position sensitive detectors (PSDs) with a 500 Hz control loop, we enhance the fiber coupling efficiency and achieve an average finite-key secure key rate of 110 kbps with an average quantum bit error rate (QBER) of 3.28 $\%$, representing a 2.5 times secure key rate enhancement under an active beam stabilization.
This work demonstrates the indispensability of active beam stabilization and provides an operational reference for robust, long-range free-space quantum communication.
\end{abstract}


\maketitle

\section{Introduction}

Quantum key distribution (QKD) utilizes the fundamental laws of quantum mechanics to allow two users (Alice and Bob) to generate a shared cryptographic key with unconditional security~\cite{Bennett1984IEEE, Ekert1991PRL, Gisin02RMP}.
While fiber-based QKD links have matured significantly, free-space QKD has recently emerged as an essential alternatives to accommodate mobile platforms (e.g., drones, vehicles, and satellites) and connect remote or inaccessible areas ~\cite{Pugh17QST, Tian24PRL, Conrad25arXiv, Li25Nature}.
Despite its potential, real-world free-space QKD is strictly limited by channel attenuation, solar background noise, and atmospheric turbulence.
Atmospheric turbulence, in particular, causes amplitude scintillation and beam wander, severely degrading the coupling efficiency of single photons into receiver detection systems ~\cite{Casado14OEng, Chen15AO, Lim23OE}.
Furthermore, utilizing practical weak coherent pulses instead of ideal single-photon sources introduces vulnerabilities to photon-number-splitting (PNS) attacks~\cite{Dusek1999OC}.
The decoy-state protocol successfully circumvents this security loophole, but its performance remains highly sensitive to channel transmission stability.

In this study, we experimentally implement a free-space QKD system in a 2 km outdoor link and thoroughly evaluate its operational performance.
To maximize the secure key rate under turbulence conditions, we present practical design optimizations for both transmitting and receiving optics, integrating high-precision active tracking and stabilization.
We quantitatively evaluate the secure key generation rate and error performance with and without active stabilization.
The realistic field data and stabilization charateristics obtained from this work provide a foundation for scaling up to free-space QKD networks, which will connect moving vehicles, personnel, and infrastructure-based stations.

\section{Experiments}
The free-space QKD system is structurally divided into two main configurations: the transmitter (Alice) and the receiver (Bob).
The transmitter comprises a multi-channel QKD light source and a transmitting optical assembly, while the receiver consists of collection optics, active stabilization system, and a QKD detection framework.

\subsection{Transmitter light source}
\begin{figure*}[t]
  \centering
  \includegraphics[width=1\textwidth]{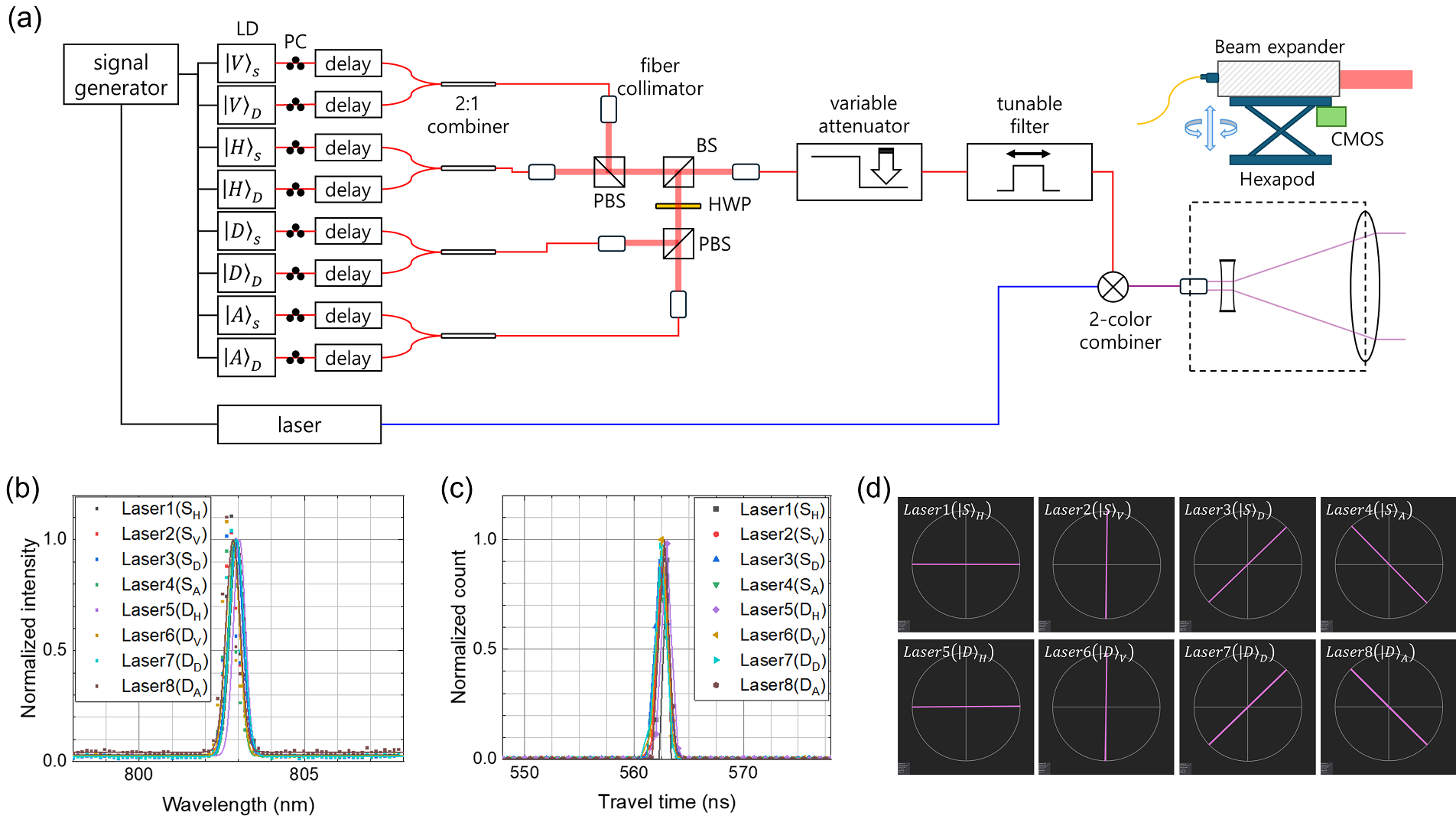}
  \caption{\label{fig1} (a) Schematic diagram of the free-space decoy-state BB84 QKD transmitter setup. 
(b) Measured optical spectra of the eight independent pulsed lasers after passing through the narrow-band tunable filter. 
(c) Temporal pulse profiles of the eight channels, demonstrating tight sub-nanosecond synchronization. 
(d) Polarization state measurements of each channel verifying the $H$, $V$, $D$, and $A$ states for both signal and decoy pulses. 
LD: laser diode, PC: polarization controller, PBS: polarization beam splitter, BS: beam splitter, HWP: half waveplate.}
\end{figure*}

The transmitter source utilizes eight independent pigtailed semiconductor lasers (Topica, iBeam smart PT-805 nm), where four units generate the signal photons and the other four produce the decoy photons, as shown in Fig.~\ref{fig1}(a).
To ensure identical pulse arrival times at the final output, individual optical delay lines are implemented in each channel to compensate for deterministic path length variations and electronic delay offsets.
The signal and decoy pulses sharing identical polarization profiles are initially merged via fiber-based 2:1 combiners.
These outputs are subsequently coupled into a single-mode fiber (SMF) using a free-space optical configuration consisting of polarization beam splitter (PBS) and beam splitter (BS).
Finally, spatially co-aligned beams are optimized via narrow-band tunable wavelength filter (WLphotonics, WLTF-NE-S) to ensure spectral indistinguishability, attenuated to the single-photon level using a variable optical attenuator, and transmitted to the receiver via a beam expander.

Each light source is a continuous-wave (CW) laser operating at a central wavelength of 805 nm, directly modulated by pulsed input voltages to generate short optical pulses.
This direct modulation inevitably induces spectral broadening and a slight shift in the peak emission wavelength, resulting in a linewidth approximately ten times broader than that of the original CW state.
Consequently, while the eight independent lasers exhibit slightly different peak wavelength after pulse modulation, their spectral envelopes overlap largely.
This overlap allows us to match both the central wavelength and the linewidth across all eight channels by implementing a narrow-band tunable filter (803 $\pm$ 0.5 nm, Fig.~\ref{fig1}(b)).

In addition to the wavelength synchronization, the effective implementation of the QKD protocol requires that all emitted laser pulses arrive at the transmitter output aperture simultaneously.
The pulse timing of each channel can differ due to variations in the lasers' internal electronic delays and individual optical path lengths.
In this work, operating at a 100 MHz repetition rate with a 2.5 ns pulse width, the pulse timing across all channels is tightly synchronized with sub-ns precision (Fig.~\ref{fig1}(c)).

To implement the decoy-state BB84 protocol, the random selection of signal, decoy vacuum, and polarization states is physically executed by feeding pseudo-random numbers into the transmitter driving module to selectively trigger the corresponding laser.
When the input random number is between 0 and 3, one of the four signal lasers is fired to produce the $\vert H \rangle$, $\vert V \rangle$, $\vert D \rangle$, or $\vert A \rangle$ state, respectively.
Random numbers from 4 to 7 trigger the corresponding decoy lasers ( $\vert H \rangle$, $\vert V \rangle$, $\vert D \rangle$, and $\vert A \rangle$).
When the random number is 8, none of the eight lasers are actuated, directly generating the vacuum state.
Consequently, the residual optical intensity during the vacuum state time slot is zero, which was experimentally verified by directly coupling the transmitter SMF output to an SPCM and observing only detector dark counts.
All optical outputs are attenuated to the single-photon level using a variable optical attenuator as depicted in Fig.~\ref{fig1}(a), and the photon number difference between the signal and decoy states is established by independently setting the driving output power of the respective laser diodes.
Furthermore, the eight lasers must maintain pre-defined polarization states at the transmitter output to support the decoy-state BB84 protocol. 
We verified these polarization states using a polarimeter (Thorlabs, PAX1000IR/M) positioned after the BS but prior to fiber coupling (Fig.~\ref{fig1}(d)).
To quantitatively substantiate state fidelity, the degree of polarization (DOP) was evaluated for each channel, as summarized in Table~\ref{table1}.
The measured DOP values across all states exceed 96 $\%$ (averaging $\sim$ 98 $\%$), confirming polarization purity and consistency between the signal and decoy states.

\begin{figure*}[t]
  \centering
  \includegraphics[width=1\textwidth]{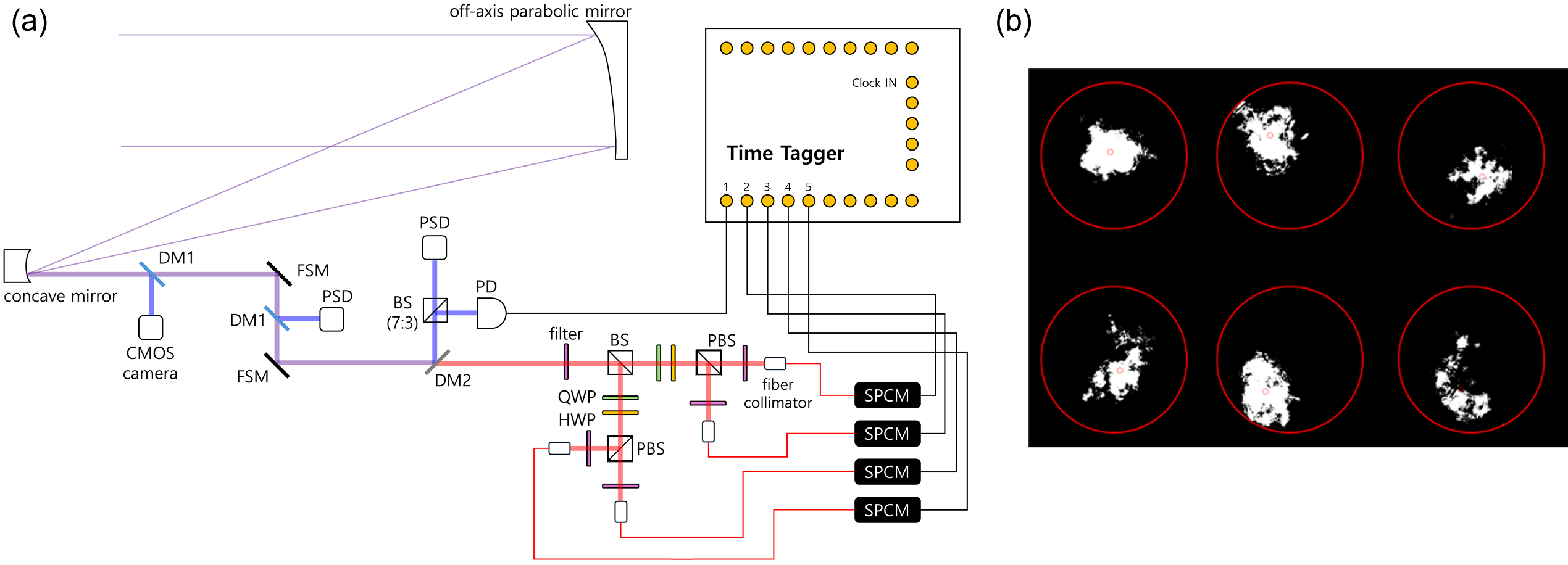}
  \caption{\label{fig2} (a) Optical layout of the free-space QKD receiver and active tracking system. 
(b) Raw intensity profiles of the 660 nm guide laser captured by the receiver CMOS camera at different time intervals. 
DM: dichroic mirror, FSM: fast-steering mirror, PD: photo-detector, PSD: position sensitive detector, BS: beam splitter, PBS: polarization beam splitter, QWP: quarter waveplate, HWP: half waveplate, SPCM: single photon counting module.}
\end{figure*}

\begin{table}[]
\caption{\label{table1} Measured degree of polarization (DOP) per channel.}
\begin{ruledtabular}
\begin{tabular}{cccc}
State & 
\begin{tabular}[c]{@{}c@{}}Signal channel DOP ($\%$)\end{tabular} & 
\begin{tabular}[c]{@{}c@{}}Decoy channel DOP ($\%$)\end{tabular} \\ 
\colrule
$\vert H \rangle$ & 98 & 99 \\
$\vert V \rangle$ & 98 & 97 \\
$\vert D \rangle$ & 98 & 99 \\
$\vert A \rangle$ & 98 & 96 \\
\end{tabular}
\end{ruledtabular}
\end{table}

\subsection{Transmitter system}
The light source for QKD (803 $\pm$ 0.5 nm) operates at an ultra-low intensity, making it highly challenging to detect using conventional optical sensors for spatial tracking.
To facilitate spatial alignment between the transmitter and receiver and actively compensate for turbulence-induced beam wander, the 803 nm photons
are co-aligned into a single optical fiber path with a 660 nm classical guide laser using a two-wavelength combiner.
The co-propagating fields are emitted via a fiber collimator and expanded through a customized beam expander consisting of a single-element concave lens and an achromatic convex doublet.
The expander is designed to ensure that both 803 nm QKD signal and 660 nm guide laser exhibit near-identical spatial diameters and divergence angles over the 2 km propagation path.
The output beam diameter ($1/e^{2}$) was measured to be 28.56 mm.

Precise alignment is essential to reliably deliver the expanded optical signals to the receiver.
Based on the Gaussian beam propagation formula $w(z)=w_{0}\sqrt{1+(\frac{z}{z_{R}})^{2}}$, where $(z_{R}=\frac{\pi w^{2}_{0}}{\lambda})$ is the Rayleigh length and $w_{0}$ is a beam waist, the beam diameter for the 803 nm wavelength expands to approximately 76 mm after propagating 2 km.
Given that the receiver's primary aperture diameter is 15 cm, the spatial margin on either side of the beam center is limited to roughly 3 cm.
Assuming a minimum alignment tolerance of 1.5 cm is required for reliable tracking, the system demands a pointing resolution of 7.5 $\mu$rad or finer.
To satisfy this requirement, we utilize a 6-axis hexapod platform (PI, H-850), which delivers a high precision within 3 - 5 $\mu$rad.
As shown in Fig.~\ref{fig1}(a), the transmitter optical assembly is integrated onto this hexapod device, where a CMOS camera with a focal length $f=50$ mm lens was also installed for coarse alignment.
At a 2 km distance, this camera provides a field of view (FOV) of 122 m $\times$ 92 m with a spatial resolution of 19 cm per pixel.
Because this pixel resolution makes pinpoint targeting challenging using the transmitter camera alone, a secondary CMOS sensor is placed on the receiver side to provide real-time spatial feedback for active position adjustment.

\begin{figure*}[t]
  \centering
  \includegraphics[width=1.0\textwidth]{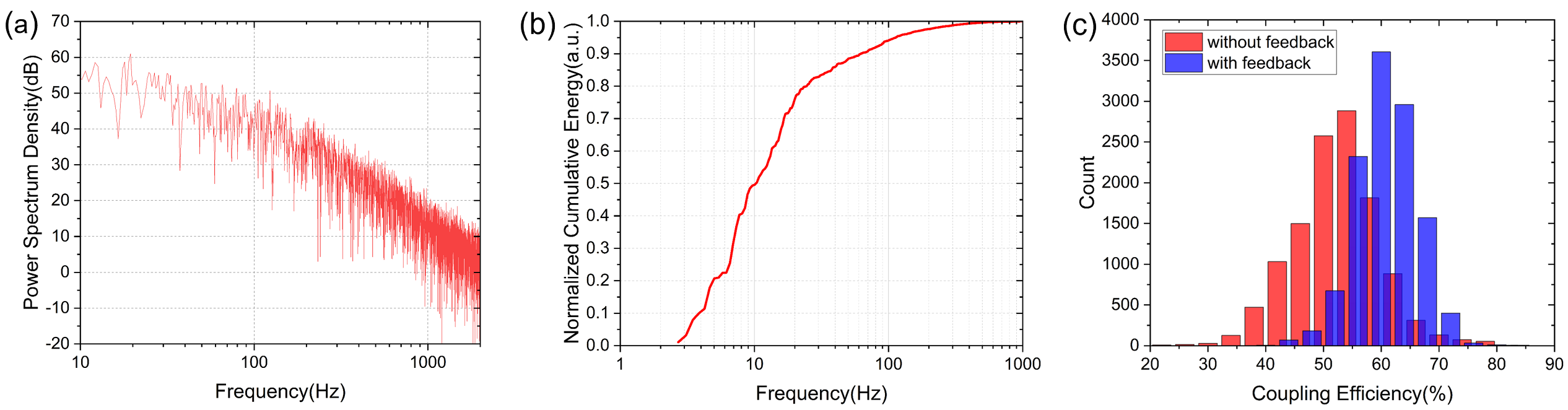}
  \caption{\label{fig3} Characterization of atmospheric turbulence and stabilization performance. 
(a) Power spectral density of the beam-wander measured by the position-sensitive detector (PSD1) at the receiver. 
(b) Normalized cumulative energy of the beam-wander as a function of frequency, showing that 90$\%$ of the turbulence energy is concentrated below 60 Hz.
(c) Histogram of the single-photon coupling efficiency into the MMF measured over a 1-minute interval, comparing the performance with (blue) and without (red) the active FSM-PSD feedback loop.}
\end{figure*}

\subsection{Receiver system and active stabilization}
As shown in Fig.~\ref{fig2}(a), the collection and measurement optics of the receiver system consists of an off-axis parabolic mirror, dichroic mirrors (DMs), fast-steering mirrors (FSMs), position-sensitive detectors (PSDs), beam splitters (BSs), polarization beam splitters (PBSs), half waveplates (HWPs), quarter waveplates (QWPs), and band-pass filters (BFs).
Initially, the co-propagating 660 nm guide laser and 803 nm photons from the transmitter enter the receiver optical assembly, where they are spatially separated by a long-pass filter (DM2).
Ideally, the 660 nm light should be entirely reflected, while the 803 nm photons are transmitted.
However, due to non-ideal component specifications, a small portion of the 660 nm guide laser inevitably leaks into the quantum channel.
Because even weak leakage can introduce significant background noise into the single-photon detectors, the transmitted fields are passed through a 658/26 nm notch filter and an 800/12 nm BF for further suppression.
Subsequently, along the both paths of the passive basis-choosing BS, the optical components are arranged in the sequence of QWP, HWP, and PBS to perform polarization state analysis.
The analyzed photons pass through an additional 800/12 nm BF before being coupled into multi-mode fibers (MMFs), which guide them to SPCMs.
The output electrical pulses are then recorded by a time-tagger module to register precise time-tagging information.
Prior to QKD implementation, polarization drift introduced by the atmospheric channel and optical components must be compensated.
To achieve this, the transmitter first sends 803 nm photons in the horizontal polarization state.
The receiver adjusts the first set of HWP and QWP until the single-photon counts on the $H$ and $V$ channels reach their maximum and minimum levels, respectively.
This process is repeated with a diagonally polarized reference beam to calibrate the second set of waveplates for the $D$ and $A$ channels.

The receiver alignment framework is divided into two primary stages: a coarse alignment, which regulates the steering direction using a 6-axis hexapod platform, and high-speed tracking, which dynamically corrects for rapid beam wander induced by atmospheric turbulence.
As shown in Fig.~\ref{fig2}(a), this stabilization system consists of a CMOS camera, two FSMs, and two PSDs.
The internal CMOS camera captures the 660 nm light split by a dichroic mirror (DM1), with its focal plane aligned to the primary off-axis parabolic mirror.
This allows the system to determine the exact spatial position of the incoming beam on the primary aperture and establish closed-loop pointing with the transmitter.
As the laser propagates through the 2 km outdoor link, it encounters atmospheric turbulence, which leads to amplitude scintillation and beam wander (Fig.~\ref{fig2}(b)).
These localized refractive index fluctuations cause the beam arrival position on the primary aperture to wander randomly in both axes.
To correct for this, the spatial position of the light source is monitored using PSD sensors, which deliver high response speeds and enable direct integration with an FPGA.
This configuration allows the system to rapidly feedback the acquired position coordinated to the FSMs for real-time beam-wander correction~\cite{Lim23OE}.

Beyond rapid turbulence, the system also experiences slow structural misalignment (long-term thermal drift) caused by the thermal expansion of mechanical components under changing environmental temperatures.
To mitigate this slow variation, a correction loop is run by the transmitter's tracking CMOS camera and the receiver's internal CMOS camera.
The system monitors the beam over a long period to track its time-averaged position.
When this positioning error drifts beyond the tracking tolerance, a calibration routine is triggered through this CMOS feedback loop to restore optimal alignment.

\begin{figure*}[t]
  \centering
  \includegraphics[width=0.9\textwidth]{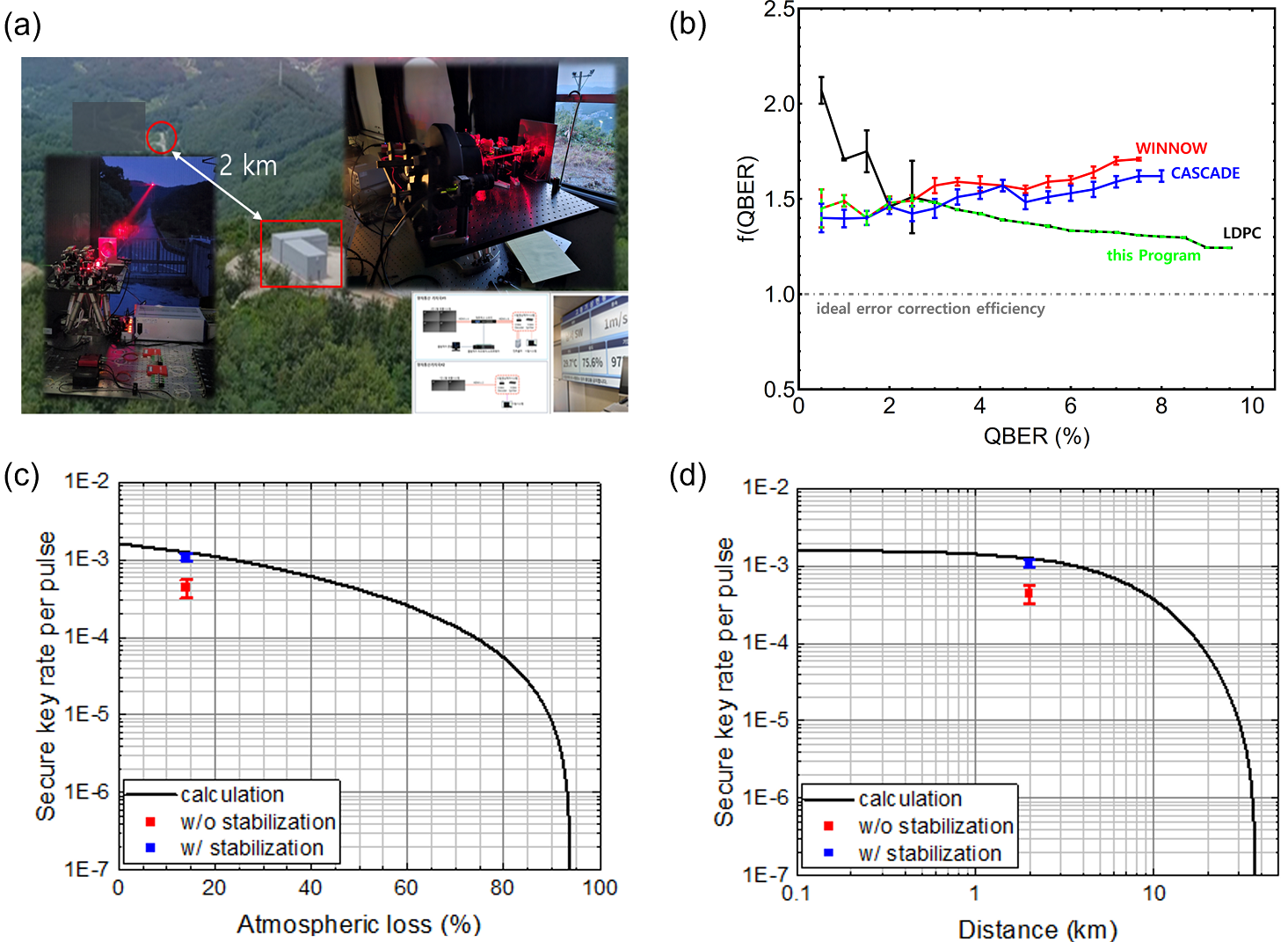}
  \caption{\label{fig4} Outdoor QKD experimental results and performance analysis. (a) Photographic overview of the 2 km free-space QKD link established at the Agency for Defense Development. 
(b) Error correction efficiency $f(\mathrm{QBER})$ as a function of the quantum bit error rate (QBER), comparing the experimental performance of the Winnow and LDPC protocols against the ideal error correction limit.
(c) Secure key rate per pulse as a function of atmospheric loss, comparing experimental data point (with and without stabilization) with the theoretical simulation.
(d) Secure key rate per pulse as a function of channel distance under clear-air conditions, showing a cutoff distance of 37 km where QBER exceeds the threshold ($\sim$ 6.5 $\%$).}
\end{figure*}

The high-speed alignment loop uses two sets of FSMs and PSDs to suppress rapid beam-wander caused by atmospheric turbulence.
To determine the frequency spectrum of the beam-wander, the position signal from PSD1 was acquired via an FPGA over a 2.6 s duration with a sampling interval of 0.2 ms (5 kHz sampling rate).
After removing the DC component (mean centering), the power spectral density of the beam center was calculated (Fig.~\ref{fig3}(a)).
The normalized cumulative energy was analyzed using $C_{k}=\frac{\sum\nolimits_{i=0}^k PSD_{i}\Delta f}{\sum\nolimits_{i=0}^N PSD_{i}\Delta f}$, where $k$ is the discrete frequency index corresponding to the target evaluation frequency $f_{k}=k\Delta f$, $N$ is the maximum frequency index corresponding to the Nyquist bandwidth ($f_{max}=N\Delta f$), and $\Delta f$ represents the frequency resolution.
As shown in Fig.~\ref{fig3}(b), approximately 90 $\%$ of the beam-wander energy is concentrated below 60 Hz.
To achieve sufficient temporal resolution and stable disturbance rejection, control systems typically require a loop rate at least 5 to 10 times higher than the target disturbance frequency ($\ge$ 300 Hz).
Here, operating the FPGA with a 40 MHz master clock, the digital control loop execution frequency was set to exactly 500 Hz via an integer clock divider to prevent internal clock jitter.

Figure~\ref{fig3}(c) shows the coupling efficiency into the MMFs measured over a 1-minute interval with and without the active stabilization.
Without feedback, the average coupling efficiency was 52.05 $\%$ $\pm$ 7.44 $\%$.
Enabling active stabilization improved the average efficiency to 61.02 $\%$ $\pm$ 5.19 $\%$, representing an absolute enhancement of approximately 8.96 $\%$p.
This increase in coupling efficiency directly improved the total system transmission from 20.68 $\%$ to 24.24 $\%$.
Consequently, the higher transmission together with the narrowed coupling variance contributed to a 17.9 $\%$ increase in the raw key rate and a 2.5 times enhancement in the finite-key secure key rate (Table~\ref{table2}), confirming the critical role of active stabilization under turbulence.

\begin{table*}[]
\caption{\label{table2} Quantitative comparison of the QKD performance parameters over the 2 km free-space link with the active beam stabilization system turned off and on.
The present values for the raw, sifted, and secure key rates, and the quantum bit error rate (QBER), represent the mathematical averages and corresponding standard deviations calculated across five independent experimental trials.}

\begin{ruledtabular}
\begin{tabular}{ccccccc}
Stabilization & 
\begin{tabular}[c]{@{}c@{}}Raw key rate\\ (kbps)\end{tabular} & 
\begin{tabular}[c]{@{}c@{}}Sifted key rate\\ (kbps)\end{tabular} & 
\begin{tabular}[c]{@{}c@{}}Secure key rate\\ (kbps)\end{tabular} & 
\begin{tabular}[c]{@{}c@{}}QBER\\ (\%)\end{tabular}  \\ 
\colrule
Off & 6,134 $\pm$ 533 & 3,113 $\pm$ 267 & 44 $\pm$ 12 & 3.59 $\pm$ 0.05 \\
On & 7,231 $\pm$ 19 & 3,570 $\pm$ 14 & 110 $\pm$ 12 & 3.28 $\pm$ 0.08 \\
\end{tabular}
\end{ruledtabular}
\end{table*}

\section{Results and Discussion}
We performed free-space QKD using the decoy-state BB84 protocol over a 2 km outdoor link at the Agency for Defense Development (ADD) (Fig.~\ref{fig4}(a)).
The experiment was conducted on clear night to avoid solar background noise during daytime.
The mean photon numbers for the signal and decoy states were configured to $\mu$ = 0.8 and $\nu$ = 0.1, respectively, with a state transmission ratio of $\mu : \nu : vac = 2 : 1: 1$.
These operating parameters were not numerically optimized prior to the trial; instead, to reliably demonstrate the operational feasibility of the system over the 2 km link, we adopted the widely utilized operating parameters reported in benchmark literature~\cite{Liao2017Nature}.
The laser sources were directly modulated at a repetition rate ($f_{\mathrm{rep}}$) of 100 MHz with a 2.5 ns pulse width.
The randomly generated optical signals from the transmitter passed through the free-space link and were captured by the SPCMs at the receiver.
A time tagger recorded the arrival sequence, channel information, and precise timestamps of the detected photons.
For the 100 MHz operation, all raw events recorded during 20 ms communication blocks were retrieved through the time tagger to perform post-processing analysis and secure key generation.

We implemented the QKD post-processing program, which includes the processes of sifting, reconciliation (error correction), and privacy amplification.
In the sifting process, only $Z$-basis signals are sifted to generate the sifted key, while the others are exploited for parameter estimation.
The reconciliation protocol operates by dividing the sifted key strings of the transmitter and receiver into blocks of a specific length and comparing the parities of these blocks to correct errors.
In this work, we employed a hybrid error correction approach that combines the Winnow~\cite{Buttler03PRA} and LDPC (low-density parity-check) codes~\cite{Gallager62IRE,MacKJay96EL} depending on the QBER.
Figure~\ref{fig4}(b) shows the error correction efficiency $f$ of our QKD post-processing program as a function of the QBER.
The error correction efficiecy $f$ represents the performance of the reconciliation process, defined as $f(\mathrm{QBER})=\frac{\text{leakEC}}{s \cdot H(\mathrm{QBER})}$, where leakEC is the number of revealed bits during the reconciliation, $s$ is the sifted key length, and $H(x)$ is the binary Shannon entropy defined as $H(x)=-x \log_{2}(x)-(1-x)\log_{2}(1-x)$.
An ideal error correction protocol achieves an efficiency of 1, whereas practical, less efficient protocols yield values exceeding 1.
In our program, the Winnow protocol was applied for QBER values below 2.65$\%$, while the LDPC code was utilized for QBER values of 2.65$\%$ and above, yielding an $f$ value between 1.2 and 1.5.
Privacy amplification is performed using a Toeplitz hash function, which is commonly used in QKD~\cite{Renner05CC}.
The output key length of the privacy amplification is calculated based on a finite-key analysis of decoy-state BB84 protocol under the non-independent and identically distributed (non-i.i.d.) assumption~\cite{Ma05PRA, Zhao06PRL,Gotteman04QIC, Devetak05PRSA, Tomamichel12NC, Curty14NC,Lim14PRA}:
\begin{equation}\label{eq1}
	l \geq s_{Z,1}^{L,KL}\left[1-H(e_{X,1}^{U,KL})\right]-\text{leakEC}-\Delta(\varepsilon)
\end{equation}
Here, $s_{Z,1}^{L}$ is the estimated lower bound of the sifted key length corresponding to the events, where both Alice and Bob used the $Z$-basis and a single photon was emitted and detected. 
$e_{X,1}^{U}$ is the estimated upper bound of the $X$-basis error rate for single-photon events, and $\Delta(\varepsilon)$ is the finite-size correction term with a maximum tolerable failure probability $\varepsilon$.
To reduce the computational complexity, we adopted the Chernoff bound with Kullback--Leibler (KL) divergence for finite-key effects~\cite{Curty14NC} rather than exploiting optimization algorithms such as semi-definite programming.

Table~\ref{table2} summarizes the QKD performance parameters of the 2 km free-space QKD link evaluated with and without active beam stabilization. 
The recorded values represent the average and corresponding standard deviation of five independent experimental trials.
Upon activating the stabilization system, the average secure key rate increased substantially from 44 kbps to 110 kbps. 
This significant performance enhancement is directly attributed to the stabilized MMF coupling efficiency facilitated by the active beam-wander correction.
Furthermore, a clear reduction in both the average QBER was observed after stabilization, demonstrating enhanced link reliability.

The secure key length $l$ was evaluated for each block of $131{,}072$ sifted key bits, with the block size determined by the dimensions of the LDPC matrix. Without stabilization, $e_{X,1}^{U,KL}=4.95 \pm 0.14$ (\%) and $s_{Z,1}^{L,KL}=64{,}408 \pm 1{,}449$. Collecting this block required $10{,}179{,}593 \pm 805{,}958$ transmission trials, and leakEC was $41{,}472$ (bits). With stabilization, the corresponding values were $e_{X,1}^{U,KL}=4.81 \pm 0.16$ (\%), $s_{Z,1}^{L,KL}=65{,}159 \pm 966$, $9{,}338{,}022 \pm 670{,}375$ transmission trials, and leakEC = $39{,}082 \pm 1{,}689$ (bits). The security parameter was set to $\varepsilon=\varepsilon_{\text{cor}}+\varepsilon_{\text{sec}}=10^{-10}$ and $\varepsilon_{\text{cor}}=\varepsilon_{\text{sec}}$ in the both cases. The correctness parameter $\varepsilon_{\text{cor}}$ bounds the probability that Alice and Bob output different keys without aborting the protocol, and the secrecy parameter $\varepsilon_{\text{sec}}$ bounds the trace distance, weighted by the probability of not aborting, between the actual key--adversary state and an ideal state in which the key is uniformly distributed an independent of all information available to the adversary.

\begin{table}[]
\caption{\label{table3} Experimental system parameters measured under operating conditions over a 2 km outdoor link.}
\begin{ruledtabular}
\begin{tabular}{cccc}
Parameter & 
\begin{tabular}[c]{@{}c@{}}Measured value \end{tabular} \\ 
\colrule
Atmospheric transmission (2 km) & 86 $\%$ \\
MMF coupling efficiency (off) & 52.053 $\pm$ 7.439 $\%$\\
MMF coupling efficiency (on) & 61.016 $\pm$ 5.188 $\%$\\
Receiver internal transmission & 77 $\%$ \\
SPCM detection efficiency ($\eta$) & 60 $\%$ \\
Total transmission ($t_{\mathrm{link}}$, off) & 20.68 $\%$ \\
Total transmission ($t_{\mathrm{link}}$, on) & 24.24 $\%$ \\
Dark count rate ($R_{\mathrm{dark}}$) & $\sim$ 1,000 cps (/channel) \\
Background photon rate ($R_{\mathrm{back}}$) & $\sim$ 7,000 cps (/channel) \\
Guide laser leakage ($R_{\mathrm{guide}}$) & $\sim$ 50,000 cps (/channel) \\
Intrinsic optical error ($e_{\mathrm{int}}$) & 0.02 (2.0 $\%$) \\
$f$(QBER) & 1.2 $\sim$ 1.5 \\
\end{tabular}
\end{ruledtabular}
\end{table}

To comprehensively characterize the 2 km outdoor link, the measured optical loss, detection efficiency, and background noise parameters under operating conditions are summarized in Table~\ref{table3}.
The single pass atmospheric transmittance over the 2 km channel was measured to be 86 $\%$ (0.66 dB loss).
The internal optical transmittance of the receiver assembly and the SPCM detection efficiency were 77 $\%$ and 60 $\%$, respectively.
Under active stabilization (on), the improved MMF coupling efficiency (61.02 $\%$) brought the total system transmission to 24.24 $\%$, compared to 20.68 $\%$ without stabilization (off).
The individual noise mechanisms contributing to the observed QBER were systematically quantified.
The accidental error count rate from each noise source is given by $R_{x}=\frac{1}{4}f_{\mathrm{rep}}P_{x}n_{\mathrm{det}}$, taking into account the basis-matching probability of 1/2 and the random error probability of 1/2.
$n_{\mathrm{det}}$ refers to the number of detectors.
Consequently, the total QBER can be analytically decomposed as~\cite{Gisin02RMP} :
\begin{equation}\label{eq2}
	\mathrm{QBER}\approx e_{\mathrm{int}}+\frac{R_{\mathrm{dark}}+R_{\mathrm{back}}+R_{\mathrm{guide}}}{R_{\mathrm{sift}}}
\end{equation}
,where $R_{\mathrm{sift}}\approx \frac{1}{2}f_{\mathrm{rep}}\mu_{\mathrm{eff}}\eta$.
Here, $\mu_{\mathrm{eff}}$ represents the effective photon number per transmitted pulse determined by the signal/decoy mean photon numbers ($\mu$, $\nu$) and their respective emission ratio ($\approx$ 0.3 in this system).
As shown in Table~\ref{table3}, the intrinsic optical error ($e_{\mathrm{int}}$ = 0.02) and residual 660 nm guide-laser leakage ($\sim$ 50,000 cps) contribute 2.0 $\%$p and 1.375 $\%$p to the QBER, respectively, while background photons (0.193 $\%$p) and detector dark count (0.028 $\%$p) have minor impacts.
The resulting analytically estimated signal QBER of 3.596 $\%$ precisely matches the experimentally observed signal QBER.

Figures~\ref{fig4}(c) and~\ref{fig4}(d) compare the simulated and experimental secure key rates per pulse as functions of atmospheric loss and channel distance, respectively.
The theoretical curves were calculated based on the decoy-state BB84 security analysis in the asymptotic limit, assuming error reconciliation efficiency $f$ = 1.2~\cite{Ma05PRA, Zhao06PRL}.
Rather than using assumed simulation inputs, all numerical parameters were replaced with values directly measured from the deployed 2 km link (summarized in Table~\ref{table3}).
In Fig.~\ref{fig4}(c), the internal optical loss, coupling efficiency, and noise rates were held fixed at their measured values while varying only the single pass atmospheric loss.
In Fig.~\ref{fig4}(d), the distance-dependent atmospheric transmittance was modeled using the Beer-Lambert law, $T_{\mathrm{atm}}(L)=\mathrm{exp}(-\gamma L)$, with the attenuation coefficient $\gamma$ = 7.541 $\times$ 10$^{-5}$ m$^{-1}$ derived from our 2 km field measurement.

Under our operating conditions, the theoretical cutoff distance is 37 km, beyond which the total QBER exceeds the security threshold ($\sim$ 6.5 $\%$), causing the secure key rate to drop to zero.
Figure~\ref{fig4}(d) shows that an operational milestone where a practical key rate of $\sim$ 1,000 bps can still be sustained at the 30 km distance with our current system parameters.
The dominant factor limiting the range beyond this cutoff is the combined effect of atmospheric attenuation and residual optical background/leakage noise (principally from the 660 nm guide laser).

To extend the reachable distance beyond 40 km, several subsystem upgrade pathways can be implemented.
First, suppressing the guide laser leakage and background noise down to the detector dark count level extends the operating range to $\sim$ 81 km (0.05 bps).
Second, upgrading the receiver hardware, by replacing the Si SPDs ($\eta$ $\sim$ 60 $\%$) with superconducting nanowire single-photon detectors (SNSPDs, $\eta$ $\sim$ 90 $\%$), optimizing the receiver internal optical transmission to 90 $\%$ (reducing intrinsic error to 0.005), and enhancing MMF coupling efficiency to 80 $\%$, extends the reachable cutoff to 51 km.

\section{Conclusion}
In this work, we have demonstrated a robust decoy-state BB84 free-space QKD system over a 2 km outdoor link under atmospheric turbulence conditions.
By combining a 100 MHz multi-channel transmitter source with high spectral and temporal overlap, high polarization purity, and a closed-loop active beam stabilization platform running at 500 Hz, we effectively reduced turbulence-induced beam wander and slow thermal drifts.
The active stabilization improved the single-photon coupling efficiency into MMFs by 8.96 $\%$p, yielding an average secure key rate of 110 $\pm$ 12 kbps under an average QBER of 3.28 $\pm$ 0.08 $\%$, 2.5 times enhancement in secure key extraction alongside a significant suppression in performance fluctuation.

Furthermore, comprehensive link modeling calibrated by our field measurements confirms an operating cutoff distance of 37 km, where secret key generation terminates as the QBER reaches the critical threshold ($\sim$ 6.5 $\%$).
Systematic scaling analyses demonstrate that extending the link distance beyond 80 km is feasible by suppressingguide laser leakage down to the dark count level, and upgrading receiver hardware to high-efficiency SNSPDs.
These results validate active spatial tracking as an essential capability for free-space quantum channels and establish a solid experimental foundation for deploying mobile terminals and long-range free-space quantum networks.

\begin{acknowledgments}
  This work was supported by an Agency for Defense Development grant funded by the Defense Acquisition Program Administration.
\end{acknowledgments}

\end{document}